\documentclass[10pt,twocolumn,twoside]{IEEEtran}  

\IEEEoverridecommandlockouts                              

\usepackage{graphics} 
\usepackage{epsfig} 
\usepackage{mathptmx} 
\usepackage{times} 
\usepackage{amsmath,amssymb,amsfonts,amsthm} 
\usepackage{mathtools}
\usepackage{optidef}
\usepackage[hidelinks]{hyperref}
\usepackage{float}
\usepackage{subfigure}
\usepackage{booktabs} 
\usepackage[table]{xcolor} 

\usepackage[moderate,title]{savetrees} 

\usepackage[skip=2pt]{caption}

\newtheorem{lemma}{Lemma}
\newtheorem{proposition}{Proposition}
\newtheorem{definition}{Definition}

\def\showcolors{0} 
\if\showcolors1
    \newcommand{\changedAlp}[1]{{\color{blue}{#1}}}
    \newcommand{\changedNick}[1]{{\color{teal}{#1}}}
    \newcommand{\changedSB}[1]{{\color{magenta}{#1}}}
    \newcommand{\changedRBlum}[1]{{\color{cyan}{#1}}}
    \newcommand{\editsNick}[1]{{\color{orange}{#1}}}
\else
    \newcommand{\changedAlp}[1]{{#1}}
    \newcommand{\changedNick}[1]{{#1}}
    \newcommand{\changedSB}[1]{{#1}}
    \newcommand{\changedRBlum}[1]{{#1}}
    \newcommand{\editsNick}[1]{{#1}}
\fi

\newcommand{\revAS}[1]{{#1}} 

\newcommand{\mainmark}[1]{{#1}}
\newcommand{\auxmark}[1]{{\widetilde{#1}}}
\newcommand{\forcingomega}{\nu}

\DeclareMathOperator{\Tr}{Tr}

\title{\LARGE \bf
Spectrum Optimization of Dynamic Networks for\\[0.2em] Reduction of Vulnerability Against Adversarial Resonance Attacks
\vspace{-0.1em} 
\thanks{\hspace{-1em}This work was supported by AFOSR award number FA9550-23-1-0046.}\\~
}

\author{Alp Sahin$^{1}$, Nicolas Kozachuk$^{2}$, Rick S. Blum$^{2}$ ~and~ Subhrajit Bhattacharya$^{1}$ \vspace{-5em}
\thanks{$^{1}$Department of Mechanical Engineering and Mechanics, Lehigh University, 19 Memorial Drive West, Bethlehem, PA 18015, U.S.A., \texttt{[als421,sub216]@lehigh.edu}.}
\thanks{$^{2}$Department of \changedRBlum{Electrical and Computer Engineering}, Lehigh University, 19 Memorial Drive West, Bethlehem, PA 18015, U.S.A., \texttt{[ngk324,rb0f]@lehigh.edu}.}
}

\date{}

\begin{document}

\setlength{\abovedisplayskip}{3pt}
\setlength{\belowdisplayskip}{3pt}

\setlength{\textfloatsep}{1pt }
\setlength{\abovecaptionskip}{1pt} 
\setlength{\belowcaptionskip}{3pt} 

\thispagestyle{empty}
\pagestyle{empty}

\maketitle

\begin{abstract}
\changedAlp{Resonance is a well-known phenomenon that happens in systems with second order dynamics. In this paper we address the fundamental question of making a network robust to signal being periodically pumped into it at \changedSB{or near} a resonant frequency by an adversarial agent with the aim of saturating the network with the signal. Towards this goal, we develop the notion of \emph{network vulnerability}, which is measured by the expected resonance amplitude on the network under a stochastically modeled adversarial attack. Assuming a second order dynamics model based on the network graph Laplacian matrix and a known stochastic model for the adversarial attack, we propose two methods for minimizing the network vulnerability 
through optimization of the spectrum of the network graph. We provide extensive numerical results analyzing the effects of both methods.}
\end{abstract}




\section{Introduction}\label{sec:introduction}
\changedSB{In this paper we consider the phenomenon of runaway amplification of signal in a network due to \emph{resonance}, which has implications on security of the network.
This is possible if an adversarial agent
pumps signal into one or more vertices of the network in a periodic manner at a frequency that matches or is very close to one of the \emph{natural frequencies} of the network. This phenomenon is observed in networks with a second order signal dynamics. 

While second order dynamics over networks has been studied in the past~\cite{doi:10.1137/110840091,1104055,6669400,7874139}, especially in context of power grids (since power transmission using alternating currents are described naturally using second-order dynamics), existing literature does not focus on controlling network parameter\changedRBlum{s} and topology for the purpose of mitigation of resonance.}

\changedAlp{We consider networks whose dynamics are governed by second order differential equations where the coefficients are functions of the graph Laplacian matrix. Assuming an adversarial signal source that obeys a known stochastic model, we develop two methods (Network Graph Optimization and Auxiliary Graph Optimization) for optimizing the network structure to reduce signal resonance under the following conditions respectively: (i) network structure can be altered by modifying the edge weights (representing the connection strength between two network nodes), (ii) edge weights of the network cannot be modified directly, but an auxiliary network can be attached to the system.}
\changedAlp{The contributions of this paper are as follows:
\begin{itemize}
    \item We provide a second-order dynamics model for signal transmission over a network under external forcing (source of the adversarial signal), that is consistent with the network topology (Section~\ref{sec:problem_description}).
    \item We develop the notion of network vulnerability, measured by the expected resonance amplitude under stochastically modeled adversarial forcing (Section~\ref{sec:problem_description}).
    \item We propose two methods, namely Network Graph Optimization and Auxiliary Graph Optimization, both of which rely on the principle of graph spectrum optimization 
    (Sections~\ref{sec:network_graph_optimization} and~\ref{sec:auxiliary_graph_optimization}).
    \item We analyze the performance of both methods through numerical experiments (Section~\ref{sec:results}).
\end{itemize}
}

\section{Related Work}\label{sec:related_work}
\changedSB{
The Laplacian dynamics on a graph, $\dot{\mathbf{x}} = -L \mathbf{x}$, as a linear signal transmission model is a model for transmission that represents \emph{diffusion} across the network and \changedRBlum{occurs} in applications frequently~\cite{mirzaev2013laplacian,7798380}. In particular, if $x_i$ is the signal value on $i$-th vertex, then this dynamics corresponds to its rate of change as a sum of the influx of the signals from its neighbors (scaled with the corresponding edge weights), minus the outflux to its neighbors. 

While first-order signal dynamics is most well-studied in context of networks~\cite{mirzaev2013laplacian,1333204,1470239}, higher-order dynamics has also been studied. 
A second-order dynamics over a network is relevant, for example, in context of distributed power grids, electrical circuits and consensus in such networks~\cite{6669400,8347206,nagpal2022designing},
where the dynamics of alternating electrical current and voltage \changedRBlum{are} naturally second order. 
The motion dynamics of mobile agents (e.g., robots) is often governed by Newtonian dynamics, which gives rise to second-order dynamics over a network of such agents~\cite{1605401}.
Second order dynamics 
can also be used to model transmission of information on social networks where the transmissibility of a signal depends both on its amount (how widespread it is) and its rate of change (how \emph{``viral''} it is). 
The properties of second-order dynamics over networks have been well-studied in \changedRBlum{the} literature (see~\cite{doi:10.1137/110840091,1104055} for example), and model reduction in the context of such dynamics has been investigated~\cite{6669400,7874139}.
However, existing literature does not focus on active control of network parameters and topology for the purpose of prevention of resonance.

Optimization of the spectrum of the Laplacian matrix in order to affect the connectivity of a network has also been extensively studied~\cite{4177054,zhang2020inter,sun2018weighted}.
However, most often, such optimization problems focus on the network connectivity in general, without explicitly addressing performance of a second-order signal dynamics over the network.

In this paper we consider a general second-order dynamics over a network with external forcing.
We particularly 
focus on developing \changedAlp{methods} for mitigating resonance attacks inflicted by an adversarial agent pumping oscillatory signal in a periodic manner at one or more vertices while trying to match a natural frequency of the network.
To our knowledge, there has been no prior work on control of resonance in a general graphical network with a focus on increasing robustness of the network to adversarial attacks.}

\section{Motivation \& \changedSB{Background}}\label{sec:problem_description}
\changedAlp{
We consider a network (\changedSB{referred to as the} \emph{main network}) represented by a weighted undirected graph \changedSB{$\mainmark{G} = (\mainmark{V}, \mainmark{E}, \mathbf{w})$ 
where $V$ is the vertex set, $E\subseteq V\times_{\mathrm{sym}} V$ is the edge set, and $\mathbf{w}$ is a set of real weights on the edges.
The vertices are indexed by natural numbers, $1,2,\cdots, n$ (where $n$ is the number of vertices), and the set of neighbors of the $k$-th vertex is denoted $\mathcal{N}_k = \{j \,|\, (k,j)\in\mainmark{E}\}$. The weight on an edge $(j,k)\in E$ is denoted by $w_{jk}$.
We also assign a natural number indexing to the edges, $1,2,\cdots,m$ (where $m$ is the number of edges), and with a little abuse of notation, $w_l$ will refer to the weight on the $l$-th edge.

The signal on the $k$-th vertex is modeled as a complex number, $x_k \in \mathbb{C}$ (while in practice the signal may be real, in which case the real part of the signal and dynamics equations are of relevance, the equations and their general solutions are most compactly represented by a complex dynamics), which follows a second order linear dynamics coupled with the signals on the neighbors of the $k$-th vertex in $\changedSB{G}$.

In it's simplest form, such a dynamics can be constructed as a natural extension of the first-order Laplacian dynamics, such that the second derivative of the signal on the $k$-th vertex is equal to the rate of \emph{influx} of signal from the neighbors of the vertex minus the rate of \emph{outflux} of signal to the neighbors, with the influx and outflux being proportional to the signal on the respective vertices. With the edge weights identified as the proportionality constants, this simple dynamics can be written as $\ddot{x}_k = \sum_{j\in \mathcal{N}_k} w_{jk} x_j - \sum_{j\in \mathcal{N}_k} w_{jk} x_k$.
This dynamics can be compactly written as $\ddot{\mathbf{x}} + L \mathbf{x} = 0$, where $\mathbf{x} \in \mathbb{\changedSB{C}}^n$ is the \textit{signal vector} (the $k$-th element of which is $x_k$) and $L = D - A$ is the weighted graph Laplacian matrix ($A$ is the \emph{weighted adjacency matrix} and $D$ the \emph{weighted degree matrix}).
The Laplacian matrix satisfies the property that its $(j,k)$-th element is zero if there does not exists an edge connecting vertices $k$ and $j$.
This property of the Laplacian matrix ensures that the dynamics of signal at a vertex depends on the signals on the neighboring vertices only, and will be referred to as the property of being \emph{consistent with the network topology}.}

\changedSB{In this paper we consider a more general form of second-order linear} dynamics \changedSB{\changedRBlum{for signals}}
following second order differential equation~\cite{meirovitch2010fundamentals}:
\begin{equation} \label{eq:network_dynamics}
    \ddot{\mathbf{x}} ~+~ \Gamma \,\dot{\mathbf{x}} ~+~ 
    \changedSB{K}
    \,\mathbf{x} ~~=~~ \mathbf{f} \,e^{i\forcingomega t}
\end{equation}
where,
\changedSB{$K$ and $\Gamma$ are called the \emph{stiffness} and \emph{damping} matrices respectively that are consistent with the network topology (\emph{i.e.}, their $(k,j)$-th element is nonzero only if there exists an edge between the $k$-th and $j$-th vertices in the graph). 
The network is subject to an adversarial \textit{forcing vector} $\mathbf{f}$ (with its $k$-th element, $f_k$, being the amplitude of adversarial signal forced on the $k$-th vertex) and \textit{forcing frequency} $\forcingomega$.

The solution to \eqref{eq:network_dynamics}, when there is no external forcing (\emph{i.e.}, $\mathbf{f} = \mathbf{0}$), exhibits oscillatory nature when the damping matrix is positive definite and the damping is small~\cite{meirovitch2010fundamentals}.
In line with the dynamics of \changedRBlum{a} signal at a vertex being the signed sum of influx and outflux of signals weighed by edge weights, we choose the stiffness matrix to be $K = L + \epsilon I$. 
The role of the $\epsilon I$ term, for a small $\epsilon>0$, is to ensure that $K$ is positive definite (all eigenvalue of $K$ are strictly greater than zero), which in turn prevent\changedRBlum{s} drift in the dynamics, since it is well-known that the weighted graph Laplacian, $L$, has a non-trivial nullspace~\cite{godsil2001algebraic}.
For notational convenience, we also define the matrix $\Omega$ such that $\Omega^2 = K = L + \epsilon I$.
\changedRBlum{We choose the damping matrix as $\Gamma = 2 \gamma \Omega^2$ for some small real $\gamma>0$, which corresponds to the fact that the damping over an edge is proportional to the edge weight (scaled by a factor of $2\gamma$). This makes both $K$ and $\Gamma$ consistent with the network topology. In the later sections, we will assume the damping multiplier $\gamma$ to be small.}
Using these new notations, we can write the dynamics \eqref{eq:network_dynamics} as
\begin{equation} \label{eq:network_dynamics_complete}
    \ddot{\mathbf{x}} ~+~ 2\,\gamma\,\Omega^2 \,\dot{\mathbf{x}} ~+~ 
    \changedSB{\Omega^2}
    \,\mathbf{x} ~~=~~ \mathbf{f} \,e^{i\forcingomega t}
\end{equation}
}
%

\begin{figure}[t]
    \centering
    \includegraphics[width=0.6\linewidth]{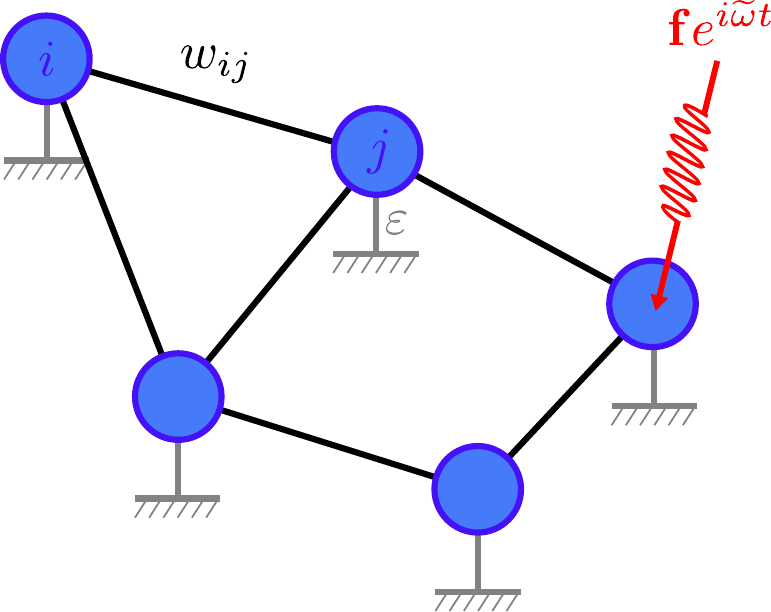}
    \caption{\changedAlp{Illustration of a network being attacked by an adversarial agent trying to cause resonance.}}
    \label{fig:graph_original}
\end{figure}


The steady-state solution to equation~\eqref{eq:network_dynamics_complete} \changedSB{is given by~\cite{meirovitch2010fundamentals}}:
\begin{equation} \label{eq:ss_network}
    \mathbf{x}_s ~=~ (-\forcingomega^2 I +  2 i \forcingomega \gamma \Omega^2 + \Omega^2)^{-1} \, \mathbf{f} ~e^{i \forcingomega t}
\end{equation}

It is a well-known fact that if the forcing frequency $\forcingomega$ matches one of the natural frequencies of the network (one of the eigenvalues of $\Omega$),
\changedSB{that leads to resonance, where,}
with a small damping, the steady-state amplitude of the forced oscillations can get arbitrarily large.
\changedSB{The objective of this paper is to minimize the expected steady-state amplitude under a probabilistic model for the distribution of the forcing frequency $\forcingomega$.}

\changedSB{We assume} that the adversarial \changedSB{agent tries to match its forcing frequency, $\forcingomega$, with one of the natural frequencies of the system (one of the eigenvalues of $\Omega$), but, is subject to uncertainties, either due to an inability to precisely select the forcing frequency, or because of an imprecise knowledge of the natural frequencies of the system. In particular, we assume that $\forcingomega$ is a stochastic variable with a probability density function \changedRBlum{dependent upon} the natural frequencies of the system.}
\begin{definition}[Network Vulnerability to Adversarial Resonance Attack]
    We define the network vulnerability to adversarial resonance attack to be
    the expected value of the squared 2-norm of the steady-state response, denoted as
    $\mathbb{E}_{\changedSB{\forcingomega,\!\mathbf{f}}} \left( \|\mathbf{x}_s\|^2_2 \right)$
\end{definition}


The main objective of this work is to develop approaches for optimization of the spectrum of the network graph (\emph{i.e.}, the spectrum of the Laplacian matrix, or equivalently, the spectrum of $\Omega^2$) to reduce the vulnerability of the network against adversarial resonance attacks with a known stochastic model.
\changedSB{We approach this problem in two different ways:
\begin{enumerate}
    \item[(1)] A direct optimization of the weights on the edges of the network that minimizes $\mathbb{E}_{\changedSB{\forcingomega,\!\mathbf{f}}} \left( \|\mathbf{x}_s\|^2_2 \right)$. We refer to this approach as \emph{Network Graph Optimization} (Section~\ref{sec:network_graph_optimization}).
    \item[(2)] When it is not possible to alter the weights on the edges directly, we propose to attach an \emph{auxiliary network} to the main network, and tune/optimize it such that this auxiliary network can effectively absorb and dissipate the excess energy from the resonance in the main network while minimizing the expected steady-state amplitude on the main network. We refer to this approach as \emph{Auxiliary Graph Optimization} (Section~\ref{sec:auxiliary_graph_optimization}).
\end{enumerate}
}

}

\section{Network Graph Optimization}\label{sec:network_graph_optimization}
Given an initial configuration of the main network specified via the graph $\mainmark{G}$, the Network Graph Optimization, refers to the procedure of optimizing the main network graph's weights and/or topology in such a way that
\changedRBlum{the vulnerability of the network is minimized against the adversarial agent's forcing behavior (forcing vector and frequency) obeying the stochastic model that will be explained in 
\revAS{Section~\ref{sec:stoch_model}.}}

\revAS{In this section, we formulate the spectrum optimization problem to minimize the vulnerability of the network (i.e., the expected value of the squared 2-norm of the steady state response).}


\subsection{\revAS{Stochastic Model of the Adversarial Forcing}}\label{sec:stoch_model}

\changedSB{We assume} that the forcing vector $\mathbf{f}$ is sampled from a uniform distribution over a $(n-1)$-unit sphere.

\begin{figure}[t]
    \centering
    \includegraphics[width=\linewidth]{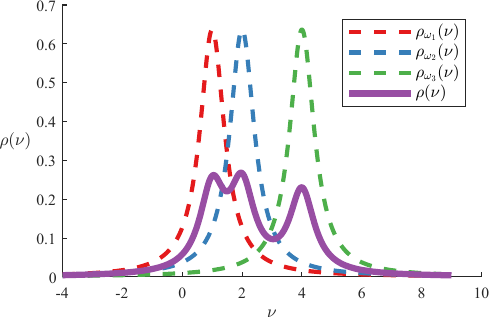}
    \caption{\changedAlp{Cauchy distributions centered at the natural frequencies $\omega_1=1$, $\omega_2=2$, and $\omega_3 = 4$ with a spread of $h=0.5$. The probability density function $\rho(\forcingomega)= \frac{1}{3} \sum\limits_{i=1}^3 \rho_{\omega_i}(\nu)$ for the adversarial agent's choice of forcing frequency is obtained as the uniformly weighted sum of the Cauchy distributions each of which are centered at the natural frequencies of the network.}}
    \label{fig:cauchy-dist}
\end{figure}

\revAS{
We assume that the adversarial agent has uncertain knowledge of the network (or equivalently precise knowledge of the network, but uncertainty/error in choosing a forcing frequency). This uncertainty/error manifests itself when the adversarial agent tries to pick a forcing frequency that matches one of the natural frequencies of the network.
%
We model this uncertainty by considering $\forcingomega$ to be a random variable whose
%
%
probability density function, $\rho$, is a uniformly weighted sum of multiple Cauchy distributions~\cite{riley2006mathematical}, each of which are centered at the natural frequencies, $\{\omega_j\}_{j=1,\dots,n}$, with a constant spread of $h$:
\begin{equation} \label{eq:pdf_omega}
    \rho(\forcingomega) = \frac{1}{n} \sum\limits_{j=1}^n \rho_{\omega_j}(\forcingomega) =\frac{1}{n} \sum\limits_{j=1}^n \frac{h/\pi}{(\omega_j - \forcingomega)^2 + h^2}
\end{equation}

The Cauchy distribution, as opposed to other probability distributions, allows the integral representing the expected value of $\|\mathbf{x}_s\|^2_2$ 
to be efficiently computed.
Figure~\ref{fig:cauchy-dist} illustrates an example where three individual Cauchy distributions are summed up with uniform weights to obtain a composite probability distribution $\rho(\forcingomega)$.
}

\subsection{\revAS{Network Vulnerability}} \label{sec:network_vulnerability}
\revAS{Following proposition computes the network vulnerability in terms of the spectrum of the network.}

    


\revAS{
\begin{proposition}[Network vulnerability] \label{prop:network_vulnerability}
If $\gamma << h$, then the network vulnerability (i.e., the expected value of the 2-norm of the steady state amplitude) is given by:
\begin{equation*}
    \mathbb{E}_{\mathbf{f},\forcingomega}\left( \|\mathbf{x}_s\|^2_2 \right) = \frac{h}{2\gamma n^2} \sum\limits_{k,j} \frac{h^2 + \omega_k^2 + \omega_j^2}{\omega_k^4 \left(h^4 + 2h^2(\omega_k^2+\omega_j^2) + (\omega_k^2 - \omega_j^2)^2\right)}
\end{equation*}
where $\omega_k$ and $\omega_j$ are the eigenvalues of $\Omega$.
\end{proposition}
}

In order to prove this result we need the following lemmas:
\changedSB{
\begin{lemma} \label{lemma:expected-value}
    If $\mathbf{f}\in\mathbb{R}^n$ is sampled from an uniform distribution over a $(n-1)$-unit sphere and $M$ is a symmetric matrix, then \[\mathbb{E}_{\mathbf{f}}(\|M \mathbf{f}\|_2^2) = \frac{1}{n} \|M\|_F^2 \]
    where $\|\cdot\|_F$ is the Frobenius norm.
\end{lemma}}
\noindent
The proof of the above lemma is deferred to Appendix~\ref{ap:lemma1} for better readability.

\begin{lemma} \label{lemma:frobenius-identity}
If $M_1$ and $M_2$ are real symmetric matrices that commute, then
$\|(M_1 + iM_2)^{-1}\|_F^2 
= \sum_{j=1}^n \frac{1}{\lambda_j(M_1)^2 + \lambda_j(M_2)^2}$, where $\lambda_j(M)$ denotes the $j$-th eigenvalue of $M$.
\end{lemma}
\noindent
The above lemma follows from the definition of the Frobenius norm, $\|M\|_F = \sqrt{\Tr(M^* M)}$ (where $M^*$ denotes the conjugate transpose of $M$).

\revAS{
\begin{proof}\emph{of Proposition~\ref{prop:network_vulnerability}:}
The expected value of $\|\mathbf{x}_s\|^2_2$ with respect to the random variables $\mathbf{f}$ and $\forcingomega$ is calculated as follows:
\begin{equation} \label{eq:exp_val}
    \mathbb{E}_{\mathbf{f},\forcingomega}\left( \|\mathbf{x}_s\|^2_2 \right) = \int\limits_{-\infty}^{\infty} \mathbb{E}_{\mathbf{f}}\left(\|\mathbf{x}_s\|^2_2\right) \,\rho(\forcingomega) ~ d\forcingomega
\end{equation}
From Lemma~\ref{lemma:expected-value} and \ref{lemma:frobenius-identity}, we have:
\begin{equation} \label{eq:ev_f0}
\begin{aligned}
    \mathbb{E}_{\mathbf{f}} \left(\|\mathbf{x}_s\|^2_2 \right) &~=~ \frac{1}{n}\|(-\forcingomega^2 I + i 2\forcingomega \gamma\, \Omega^2 + \Omega^2)^{-1}\|^2_F \\
    & = \frac{1}{n}\sum\limits_k \frac{1}{(\omega_k^2 - \forcingomega^2)^2 + (2 \gamma\, \forcingomega\omega_k^2)^2}
\end{aligned}
\end{equation}
Substituting 
equation~\eqref{eq:ev_f0} into equation~\eqref{eq:exp_val},
\begin{equation} \label{eq:ev_omega}
    \mathbb{E}_{\mathbf{f},\forcingomega}\left( \|\mathbf{x}_s\|^2_2 \right) = \frac{h}{\pi n^2} \sum\limits_{k,j} g(\omega_k^2,\omega_j^2)
\end{equation}
where
\begin{equation} \label{eq:g-interation}
g(\omega_k^2,\omega_j^2) = \int\limits_{-\infty}^{\infty} \frac{d\forcingomega}{((\omega_k^2 - \forcingomega^2)^2 + (2 \gamma\forcingomega\omega_k^2)^2)((\omega_j - \forcingomega)^2 + h^2)}    
\end{equation}
\changedSB{Since $\gamma$ is non-zero, the poles of the integrand above lie away from the real line on the complex plane, and hence}
a closed-form expression for the integral $g(\omega_k^2,\omega_j^2)$ can be obtained using the Residue theorem~\cite{saff2013fundamentals} by 
\changedSB{performing a contour integration over the real line and} a semi-circular 
\changedSB{arc of radius $R\to\infty$ on the upper half of the complex plane (Figure~\ref{fig:integration-contour})}.
%

\begin{figure}[t!]
  \centering
  \includegraphics[width=\linewidth,trim=0 13 0 0,clip=true]{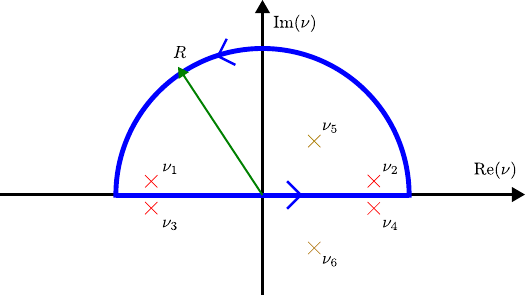}
  \caption{\changedAlp{Integration contour for equation~\eqref{eq:g-interation}. Poles $\forcingomega_1$ to $\forcingomega_4$ correspond to the forcing vector component $\mathbb{E}_{\mathbf{f}} \left(\|\mathbf{x}_s\|^2_2 \right)$ and they collapse on to the real line as $\gamma$ goes to zero. Poles $\forcingomega_5$ and $\forcingomega_6$ correspond to the forcing frequency component $\rho(\forcingomega)$.}}
  \label{fig:integration-contour}
\end{figure}
Assuming $\gamma \ll h$,
we can compute the roots of the quartic polynomial in $\forcingomega$ in the denominator of the integrand in \eqref{eq:g-interation} using a symbolic algebra toolbox, and then apply the Residue theorem to obtain

\begin{equation}
    g(\omega_k^2,\omega_j^2) = \frac{\pi}{2\gamma} \frac{h^2 + \omega_k^2 + \omega_j^2}{\omega_k^4 \left(h^4 + 2h^2(\omega_k^2+\omega_j^2) + (\omega_k^2 - \omega_j^2)^2\right)}
\end{equation}
This proves the proposition.
\end{proof}
}
\changedSB{The objective is to minimize this expected value of the $2$-norm of the steady-state amplitude, so as to mitigate the effects of resonance attacks on the network.
We note that $\omega_k$ and $\omega_j$ are the eigenvalues of $\Omega = \sqrt{L + \epsilon I}$, where the Laplacian matrix, $L=D-A$, depends on the weights on the edges of the graph. Thus $\mathbb{E}_{\mathbf{f},\forcingomega}\left( \|\mathbf{x}_s\|^2_2 \right)$, as described in \revAS{Proposition~\ref{prop:network_vulnerability}}, is a function of the edge weights of the graph. We thus define the objective function, $J(\mathbf{w}) = \mathbb{E}_{\mathbf{f},\forcingomega}\left( \|\mathbf{x}_s\|^2_2 \right)$ to be a function of the edge weight vector, $\mathbf{w}\in\mathbb{R}^m$ (where $m$ is the number of edges in the graph). It can be checked that $J$ is in general a non-convex function.}
However, if $h\rightarrow 0$, it can be indeed shown that $J$ is convex in the edge weights.
\begin{proposition}
    For a sufficiently large value of $h$, $J(\mathbf{w})$ is convex. 
\end{proposition}
\begin{proof}\emph{Sketch:}
    Define the symmetrized function $\widetilde{g}(\omega_k^2,\omega_j^2) = \frac{1}{2} \left( g(\omega_k^2,\omega_j^2) + g(\omega_j^2,\omega_k^2) \right)$ so that $J(\mathbf{w}) = \frac{h}{\pi n^2} \sum_{k,j} \widetilde{g}(\omega_k^2,\omega_j^2)$.
    
    Since $\{\omega_j^2\}_{j=1,2,\cdots,n}$ are eigenvalues of $\Omega^2 = L + \epsilon I$, we can write $J(\mathbf{w}) 
    = \frac{h}{\pi n^2} \Tr(\widetilde{g}(L + \epsilon I, L + \epsilon I))$ (where $\widetilde{g}(M,N)$ refers to the matrix extension of the scalar function, $\widetilde{g}$~\cite{2025arXiv250114515B,bhatia}).
    It is known that the trace of the matrix extension of a scalar function inherits the convexity properties of the scalar function (see our technical report \cite{2025arXiv250114515B} for a detailed proof for the case of multi-variable scalar functions), and as a consequence of that, it's sufficient to show that the function $\widetilde{g}$ is convex.

    When $h$ is sufficiently large (compared to the eigenvalues of $L$), the function $\widetilde{g}$ becomes $\widetilde{g}(x,y) = \frac{\pi}{4\gamma} \frac{h^2 + x + y}{h^4 + 2h^2(x+y) + (x - y)^2} \left(\frac{1}{x^2} + \frac{1}{y^2}\right) \simeq \frac{\pi}{4\gamma h^2} \left(\frac{1}{x^2} + \frac{1}{y^2}\right) $.
    It s easy to show that this function in convex in $\mathbb{R}_{+}^2$ (a direct computation of the Hessian shows that its eigenvalues are positive). This proves the proposition.
\end{proof}
As a consequence of the above proposition, while $J(\mathbf{w})$ may not be strictly convex for all values of $h$, 
when $h$ is large (corresponding to high uncertainty in the adversarial agent's ability to choose/apply a forcing that matches a natural frequency of the graph), the objective is indeed concvex.

\subsection{Spectrum Optimization of the Main Network Graph}
\label{sec:eigenspectrum_optimization}
We define the \emph{spectrum optimization problem of the main network graph} \changedSB{as the problem of minimizing the expected steady-state amplitude of signal on the network under the described stochastic forcing:} 
\begin{mini}|l|
{\mainmark{\mathbf{w}}}  {\mainmark{J}\changedSB{(\mathbf{w})}}
{}{}
\addConstraint{\mathbf{1}^T\mainmark{\mathbf{w}} = w^{tot}}
\addConstraint{\mainmark{\mathbf{w}} \succeq w^{min}\mathbf{1}}
\end{mini}
where $\mainmark{\mathbf{w}} \changedSB{\in\mathbb{R}^m}$ \changedSB{is the vector of} weights on the network graph edges. 
Here we treat the total \changedRBlum{sum} of weights, $\changedSB{\sum_{j=1}^m w_j =} w^{tot} \geq 
m\,
w^{min} \geq 0$, as a resource to be re-distributed among all edges, hence their sum is \changedSB{constrained to be} equal to $w^{tot}$. $w^{tot}$ is assumed to be specified by the initial weight distribution on the network graph $\mainmark{G}$.
We consider non-negative edge weights throughout \changedSB{the paper}, which further \changedRBlum{imply} $w^{min} > 0$ to preserve the connectivity \changedSB{and network topology}. Note that $\mainmark{\mathbf{w}}$ only contains the weights of the existing edges on the graph, thus it is not possible to remove existing edges or add non-existent edges during the optimization.

\begin{figure}[t]
    \centering
    \includegraphics[width=\linewidth]{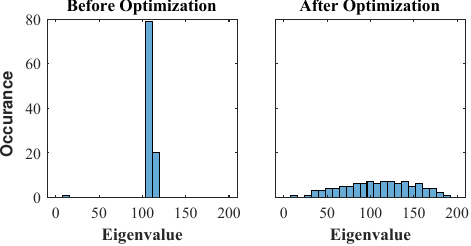}
    \caption{Histograms of the Laplacian matrix eigenvalues for the initial network graph $\mainmark{G}$ and optimized network graph $\mainmark{G}^*$. The initial network is modeled by a complete graph, whose edge weights are perturbed away from a uniform distribution by a small amount. The corresponding spectrum (on the left) is \emph{peaky}, whereas as a result of the spectrum optimization, the spectrum (on the right) has become \emph{flatter}.}
    \label{fig:eig_spec_illustration}
\end{figure}

The optimal edge weights are denoted by $\mainmark{\mathbf{w}}^*$ and the corresponding optimal \changedSB{weighted} graph is $\mainmark{G}^* \changedSB{=(V,E,\mainmark{\mathbf{w}}^*)}$.
In Figure~\ref{fig:eig_spec_illustration}, we provide a histogram of eigenvalues of the graph Laplacian matrix (henceforth referred to as the \emph{eigenvalue spectrum}) for both the initial network graph $\mainmark{G}$ and the optimized network graph $\mainmark{G}^*$, where both graphs are complete (\emph{i.e.}, there exists an edge between every pair of vertices in $V$).
As can be seen, the optimization has the effect of \emph{flattening} the eigenvalue spectrum, resulting in a more uniform distribution of the eigenvalues, compared to the initial \emph{peaky} spectrum where the eigenvalues are accumulated around a specific value.

Observing that the eigenvalues of the graph Laplacian, $\{\lambda_k\}_{k=1,2,\cdots,n}$, and the eigenvalues of $\Omega$, $\{\omega_k\}_{k=1,2,\cdots,n}$, are related monotonically as $\omega_k = \sqrt{\lambda_k + \epsilon}$, the interpretation of this change in the eigenvalue spectrum is as follows: If a graph has a peaky spectrum, an adversarial agent will have a higher chance of success in causing resonance (high-amplitude oscillations) in the graph by choosing the frequency near the peak to pump its forcing signal into the graph. Whereas, with a flattened spectrum, it has less obvious peak to choose from, and hence the overall expected steady-state amplitude is lower.

\section{Auxiliary Graph Optimization}\label{sec:auxiliary_graph_optimization}

\changedAlp{
We consider the scenario where the main network cannot be manipulated directly \changedSB{and} the edge weights of the main graph $\mainmark{G}$ cannot be modified. 
\changedSB{An alternative to changing the network itself at the level of individual edges of the network is to connect the network with an \emph{auxiliary network} that is tuned/optimized 
}
in a way that minimizes the vulnerability of the main network.
%
\changedSB{This idea of using auxiliary systems to dampen certain frequencies of oscillation appear extensively in the study and design of mechanical and structural systems (such as the use of \emph{tuned mass dampers} in prevention of mechanical vibrations in buildings~\cite{https://doi.org/10.1002/tal.1068}).
We, however, develop the mathematical foundations and \changedAlp{methods} for designing analogous tuned auxiliary networks for mitigating resonance attacks on the network by an adversarial agent.}

In this section, we first reformulate the dynamics equations and the definition of vulnerability based on the \emph{combined network} (main network $+$ auxiliary network). Then, we derive the corresponding objective function and formulate the spectrum optimization problem to minimize the vulnerability of the main network.

\subsection{Formulation of Combined Dynamics}

We denote the graph representation of the auxiliary network \changedSB{by} $\auxmark{G}$, and the combined network \changedSB{is denoted by $\mainmark{G} \cup \auxmark{G}$ (see Figure~\ref{fig:graph_aux}). A second-order unforced signal dynamics on the stand-alone auxiliary network is given by $\ddot{\widetilde{\mathbf{x}}} + \widetilde{\Gamma} \dot{\widetilde{\mathbf{x}}} + \widetilde{K} \widetilde{\mathbf{x}} = 0$, where $\widetilde{\mathbf{x}} \in \mathbb{C}^{\auxmark{n}}$ is the signal vector on the vertices of the auxiliary-network, and $\widetilde{\Gamma}$ and $\widetilde{K}$ are the damping and stiffness matrices respectively that are consistent with the topology of the auxiliary network (in particular, $\widetilde{K} = \widetilde{\Omega}^2 = \widetilde{L} + \epsilon I$ and $\widetilde{\Gamma} = 2\widetilde{\gamma}\widetilde{\Omega}^2$ (where $\widetilde{L}$ is the weighted Laplacian matrix of the auxiliary network and $\widetilde{\gamma}$ is the damping multiplier on the auxiliary network).

We make the following simplifying assumptions about the auxiliary network and its inter-connection with the main network:
\begin{itemize}
    \item[i.] We assume the auxiliary network to have the same number of vertices as the main network (that is, $\auxmark{n}=n$).
    \item[ii.] The above assumption allows a one-to-one connection between the vertices of $G$ and $\widetilde{G}$. The indexing of the vertices of $\widetilde{G}$ is done in a way that the $k$-th vertex of $G$ is assumed to be connected with (and only with) the $k$-th vertex of $\widetilde{G}$.
    \item[iii.] The inter-connecting edges between $G$ and $\widetilde{G}$ are assumed to have stiffness (corresponding to a weight of $c$ on those edges), but no damping, allowing the second derivative of the signal on a vertex in $G$ to be coupled with the signal on the neighbor in $\widetilde{G}$, but not its first derivative.
    \item[iv.] It's assumed that the adversarial agent can attack vertices of the main network, but not the auxiliary network.
    \item[v.] The connectivity of the auxiliary graph is specified via one of the two types: (1) a \emph{mirrored} auxiliary graph, which exactly mirrors the connectivity of the main graph, and (2) a \emph{complete} auxiliary graph, which is a complete graph.  Note that when the main graph is complete, both types correspond to the same auxiliary graph.
\end{itemize}}


\begin{figure}
    \centering
    \includegraphics[width=\linewidth]{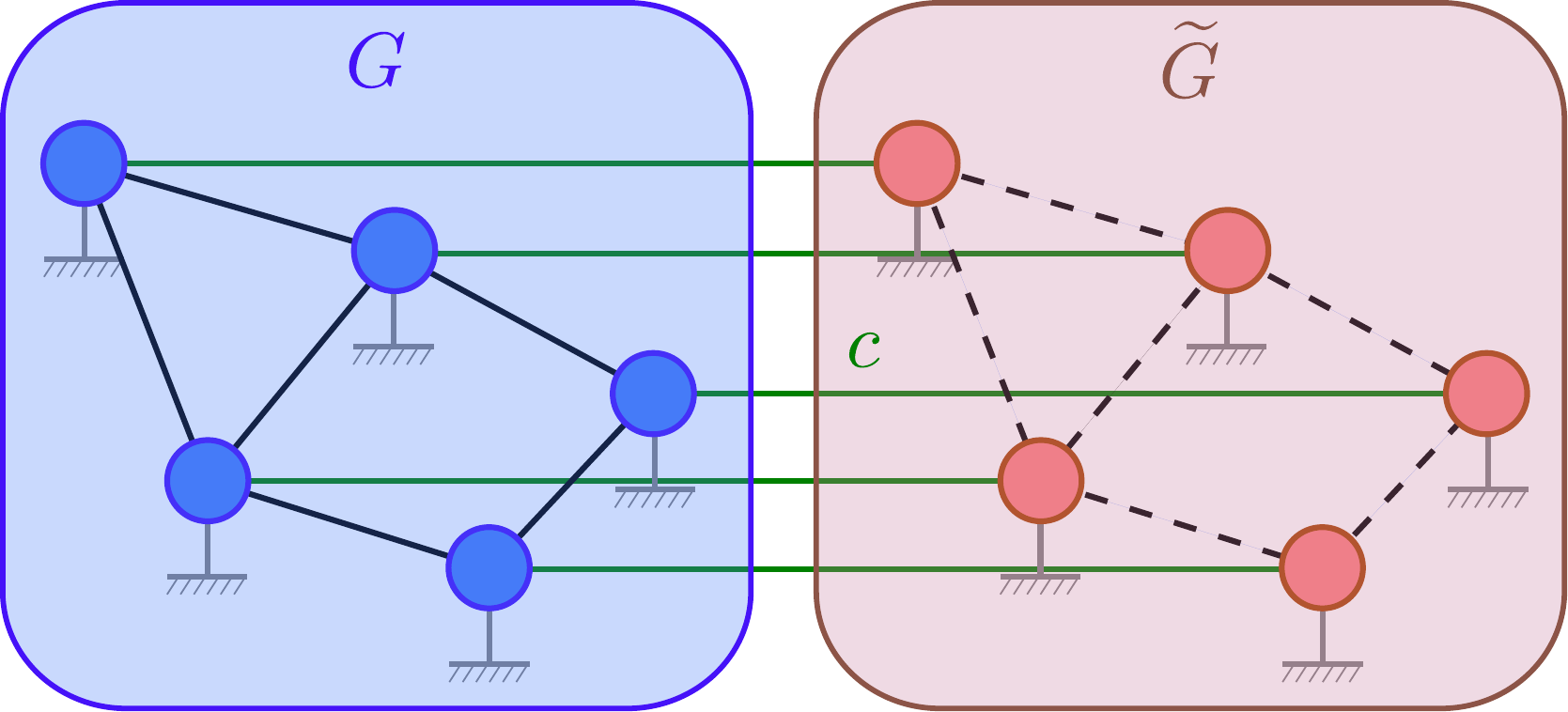}
    \caption{\changedAlp{Illustration of an auxiliary graph $\auxmark{G}$ attached to the original graph $\mainmark{G}$ with an aim to decrease vulnerability against adversarial attacks. The auxiliary graph is of type \emph{mirrored} (has the same connectivity as the main graph). Green lines indicate the inter-graph connections with weights $c$.}
    }
    \label{fig:graph_aux}
\end{figure}

\changedSB{Since} the auxiliary network is connected to the main network, with the purpose of mitigating the resonance on the main network under adversarial forcing, \changedSB{based on the above assumptions, the signal dynamics over $G$ and $\widetilde{G}$ are coupled to give the following signal dynamics on $G\cup\widetilde{G}$}: 
\begin{equation} \label{eq:aux_dynamics}
    \begin{bmatrix}
        \Ddot{\mainmark{\mathbf{x}}} \\
        \Ddot{\auxmark{\mathbf{x}}} \\
    \end{bmatrix} + \begin{bmatrix}
        \mainmark{\Gamma} & 0 \\
        0 &\auxmark{\Gamma}
    \end{bmatrix} 
    \begin{bmatrix}
        \Dot{\mainmark{\mathbf{x}}} \\
        \Dot{\auxmark{\mathbf{x}}}
    \end{bmatrix} 
    + \begin{bmatrix}
        \mainmark{K} + \changedSB{c} I & -\changedSB{c} I \\
        -\changedSB{c} I & \auxmark{K} + \changedSB{c} I
    \end{bmatrix} 
    \begin{bmatrix}
        \mainmark{\mathbf{x}} \\
        \auxmark{\mathbf{x}}
    \end{bmatrix} ~=~ 
    \begin{bmatrix}
        {\mathbf{f}} \\
        0
    \end{bmatrix}e^{i \forcingomega t} 
\end{equation}
\changedSB{where the terms $\changedSB{c} I$ represent coupling between the dynamics of the two networks due to the one-to-one connection between the vertices of $G$ and $\widetilde{G}$, and affects the stiffness matrix of the combined network, but not the damping matrix.}
An illustration of a combined network is provided in Figure~\ref{fig:graph_aux}.


\subsection{\revAS{Network Vulnerability with Attached Auxiliary Network}}
\revAS{
Following proposition gives the vulnerability of a network to which we attach the auxiliary network.
\begin{proposition}[Network vulnerability with attached auxiliary network] \label{prop:nv_aux}
    If $\mainmark{\Omega}$ and $\auxmark{\Omega}$ are simultaneously diagonalizable, then the vulnerability of a network to which an auxiliary network is attached is given by:
    \begin{equation*}
    \mathbb{E}_{\mathbf{f},\forcingomega}\left(\|\mainmark{\mathbf{x}}_s\|^2_2\right) = \frac{1}{n^2} \sum\limits_{k,j} \int\limits_{-\infty}^{\infty} s_k(\forcingomega) \overline{s}_k(\forcingomega) \rho_{\omega_j}(\forcingomega) d\forcingomega
    \end{equation*}
where $\rho_{\omega_j}(\forcingomega) = \frac{h / \pi}{(\mainmark{\omega}_{j}-\forcingomega)^2 + h^2}$, and,
\begin{equation*}    
s_k(\forcingomega) = \frac{1}{-\forcingomega^2 + i\forcingomega2\mainmark{\gamma}\mainmark{\omega}_k^2 + \mainmark{\omega}_{k}^2 + \changedSB{c}  - \frac{\changedSB{c}^2}{-\forcingomega^2 + i\forcingomega2\auxmark{\gamma}\auxmark{\omega}_{k}^2 + \auxmark{\omega}_{k}^2 + \changedSB{c}}}
\end{equation*}
with $\auxmark{\omega}_{k}$ denoting the $k$-th eigenvalue of $\auxmark{\Omega}$ and $\overline{s}_k$ denoting the complex conjugate of $s_k$.
\end{proposition}
}
\revAS{
\begin{proof}

\changedSB{The steady-state solution to \eqref{eq:aux_dynamics} is
\begin{equation} \label{eq:aux_ss}
    \begin{bmatrix}
        \mainmark{\mathbf{x}}_s \\
        \auxmark{\mathbf{x}}_s \\
    \end{bmatrix} = S^{-1} 
    \begin{bmatrix}
        \mathbf{f} \\
        0 \\
    \end{bmatrix} e^{i\forcingomega t}
\end{equation}
where,}
\begin{equation} \label{eq:S-matrix}
\small
    S = \begin{bmatrix}
        i\forcingomega2\mainmark{\gamma}\mainmark{\Omega}^2 + \mainmark{\Omega}^2 + (\changedSB{c}-\forcingomega^2) I & -\changedSB{c} I \\
        -\changedSB{c} I& i\forcingomega2\auxmark{\gamma}\auxmark{\Omega}^2 + \auxmark{\Omega}^2 + (\changedSB{c} -\forcingomega^2)I
    \end{bmatrix}
\end{equation}

\changedSB{However, we note that}
we are only interested in the response of the main network to the adversarial attacks, which 
\changedSB{from \eqref{eq:aux_ss} and \eqref{eq:S-matrix}} is:

\begin{equation} \label{eq:ss_cf_aux}
\begin{aligned}
    \mainmark{\mathbf{x}}_s &= [S^{-1}]_{11} {\mathbf{f}} e^{i\forcingomega t}
\end{aligned}
\end{equation}
where $[S^{-1}]_{11}$ is the top left \changedSB{$n\times n$} block of the inverse of the matrix $S$, \changedSB{which can be computed using Schur complement of a block matrix~\cite{boyd2004convex} as}:
\begin{equation} \label{eq:s-inv-11}
\begin{aligned} 
    [S^{-1}]_{11} & = ([S]_{11} - [S]_{12}[S]_{22}^{-1}[S]_{21})^{-1} \\
    & = \Big( \left( i\forcingomega2\mainmark{\gamma}\mainmark{\Omega}^2 + \mainmark{\Omega}^2 + (\changedSB{c}-\forcingomega^2) I \right) \\
    & \qquad +~ \changedSB{c}^2 \left( i\forcingomega2\auxmark{\gamma}\auxmark{\Omega}^2 + \auxmark{\Omega}^2 + (\changedSB{c} -\forcingomega^2)I \right)^{-1} \Big)^{-1}
\end{aligned}
\end{equation}

\changedSB{(As a quick sanity check,} note that when $\changedSB{c}=0$, which means that the main and the auxiliary networks are not connected, \changedSB{we have}
\begin{equation} \label{eq:sanity}
\begin{aligned}
    [S^{-1}]_{11} 
    &= (i\forcingomega2\mainmark{\gamma}\mainmark{\Omega}^2 + \mainmark{\Omega}^2 - \forcingomega^2 I)^{-1}
\end{aligned}
\end{equation}
\changedSB{indicating that} the steady-state response on the main network is equivalent to the one derived in equation~\eqref{eq:ss_network}, \changedSB{as expected}. In Section~\ref{sec:results} we use this theoretical result to perform \changedSB{further numerical} sanity check on the \changedSB{Auxiliary Graph Optimization} objective function.)

Since $\mainmark{\Omega}$ and $\auxmark{\Omega}$ are simultaneously diagonalizable, \changedSB{using \eqref{eq:s-inv-11},} allows us to 
%
\changedSB{compute the eigenvalues of $[S^{-1}]_{11}$ as}
\begin{equation} \label{eq:s_k}
s_k(\forcingomega) = \frac{1}{-\forcingomega^2 + i\forcingomega2\mainmark{\gamma}\mainmark{\omega}_k^2 + \mainmark{\omega}_{k}^2 + \changedSB{c}  - \frac{\changedSB{c}^2}{-\forcingomega^2 + i\forcingomega2\auxmark{\gamma}\auxmark{\omega}_{k}^2 + \auxmark{\omega}_{k}^2 + \changedSB{c}}}
\end{equation}

According to the stochastic model explained in Section~\ref{sec:stoch_model}, $\mathbf{f}$ is being uniformly sampled from ($n-1$)-unit sphere and the adversarial agent only has imprecise information about the main graph (\emph{i.e.}, it has no information about the auxiliary graph and hence the combined network) leading to the probability density function \eqref{eq:pdf_omega} for the forcing frequency $\forcingomega$.

Rest of the proof is similar to the proof of Proposition~\ref{prop:network_vulnerability}.
We use Lemma~\ref{lemma:expected-value} and \ref{lemma:frobenius-identity}, and equation~\eqref{eq:s_k} to compute the expected value with respect to $\mathbf{f}$. the result of the proposition then follows from the substitution of this expected value together with the p.d.f.~\eqref{eq:pdf_omega} into equation~\eqref{eq:exp_val}.
\end{proof}

\changedAlp{Later on, we will show that there will be an approximation error between the computed expected value and the average squared $2$-norm of the steady-state response when $\mainmark{\Omega}$ and $\auxmark{\Omega}$ are not simultaneously diagonalizable.}
}

%
A closed form expression for the integral \revAS{in Proposition~\ref{prop:nv_aux}} is obtained using the Residue theorem with the same contour \changedSB{as before as described in Section\revAS{~\ref{sec:network_vulnerability}}.
In order to use the Residue theorem as described, however, one needs to compute the roots of the quartic polynomial in $\forcingomega$ in the denominator of the integrand and determine whether those roots have positive or negative imaginary parts. In this case a direct computation of that, even using a symbolic algebra toolbox, was not feasible because of the complexity of the problem. In order to simplify computation of the roots, we use linearization with respect to $\gamma$. The details of the computation are provided in Appendix~\ref{ap:root}.}
Corresponding calculations are performed using a symbolic mathematics toolbox. We omit the resulting expression for brevity.

Assuming that the main graph $\mainmark{G}$ and the parameters $n$, $h$, $\mainmark{\gamma}$ remain constant,
\changedSB{the objective is to minimize $\mathbb{E}_{\mathbf{f},\forcingomega}\left( \|\mainmark{\mathbf{x}}_s\|^2_2 \right)$, which is a function of the eigenvalues of the auxiliary stiffness matrix $\auxmark{\Omega}$ (which, in turn, are functions of the weights on the auxiliary graph edges, $\auxmark{\mathbf{w}}$), the uniform inter-graph edge weight $\changedSB{c}$ and the auxiliary damping factor $\auxmark{\gamma}$.} 
\changedRBlum{For the purposes of this paper, we assume $\auxmark{\gamma}$ to be a small constant, \changedSB{in order to allow signals transmitted over the network (non-adversarial) to persist and not get dissipated too quickly}.} The \changedSB{resulting} objective function is \changedSB{thus} defined as:
\begin{equation}
    \auxmark{J}(\changedSB{\auxmark{\mathbf{w}}},\changedSB{c}) ~=~ \mathbb{E}_{\mathbf{f},\forcingomega}\left( \|\mainmark{\mathbf{x}}_s\|^2_2 \right)
\end{equation}

\subsection{Spectrum Optimization of the Auxiliary Network Graph}
We define the spectrum optimization problem of the auxiliary network graph as follows:
\begin{mini}|l|
{\changedRBlum{\auxmark{\mathbf{w}},c}}  {\changedRBlum{\auxmark{J}(\auxmark{\mathbf{w}},c)}}
{\label{eq:aux-optimization}}
{}
\addConstraint{\mathbf{0} \preceq \auxmark{\mathbf{w}}}
\addConstraint{0 \leq \changedSB{c}}
\addConstraint{\mathbf{1}^T\auxmark{\mathbf{w}} + n \changedSB{c} \leq r_m w^{tot}}
\end{mini}
Here, we assume that the weight resource is specified as a multiple of the total weights on the main graph (denoted by $r_m w^{tot}$) which is to be distributed among the auxiliary graph and inter-graph edges. We consider non-negative edge weights throughout, without any additional lower bound.

The optimal auxiliary graph edge weights are denoted by $\auxmark{\mathbf{w}}^*$, the optimal inter-graph edge weight is $\changedSB{c}^*$ and the corresponding optimal auxiliary graph \changedSB{weight} configuration is $\auxmark{G}^*$.

\changedRBlum{Note that it is also possible to consider the case where the auxiliary damping multiplier $\auxmark{\gamma}$ is a decision variable. We include further discussion on the effects of the auxiliary damping and experimental results in Section~\ref{sec:res-aux-damp}.}
}

\changedNick{




}

\section{Results}\label{sec:results}

\changedAlp{
In this section, we present experiments conducted to accomplish the following:
\begin{itemize}
    \item Validate the accuracy of the objective functions, $\mainmark{J}$ and $\auxmark{J}$, in representing the network vulnerability measured by $\mathbb{E}(\|\mainmark{\mathbf{x}}_s\|^2_2)$ for $\mainmark{\mathbf{x}}_s$ defined on $\mainmark{G}$ and $\auxmark{G}$, as described \revAS{in Proposition~\ref{prop:network_vulnerability} and~\ref{prop:nv_aux}} respectively.
    \item Analyze the effects of the problem parameters associated with the network dynamics and constraints on the relative vulnerability decrease that can be achieved via the proposed methods.
    \item Demonstrate the effectiveness of the proposed methods in decreasing the network vulnerability across a variety of problem instances.
    \item \changedSB{Perform numerical simulation of dynamics over a network to further validate} the results achieved by the Network Graph Optimization.
    \item \revAS{Apply the network graph optimization to the communication network among a team of mobile robots between which the signal strength decays with increasing distance.}
\end{itemize}

\subsection{Implementation Details and Setup} \label{sec:res-setup}

We solve \changedSB{the} network graph and auxiliary graph spectrum optimizations using the \emph{interior-point} algorithm
~\cite{MATLABOptimization2023}.

\subsubsection{Network Graph Construction}
\label{sec:sg_processing}

\changedSB{All algorithms are implemented and tested on three classes of network graphs: 
\begin{enumerate}
    \item[i.] \emph{Random Complete Graphs} (``RCG''): Given a desired number of vertices, $n$, we establish an edge between every pair of vertices, thus resulting in a graph with $n_e = \dim \mainmark{\mathbf{w}} = {n \choose 2}$ edges. We then sample the weight for each edge from an uniform distribution on the interval $[1-w_{p},1+w_{p}]$, where $w_{p}$ is a given \emph{weight perturbation}.
    \item[ii] \emph{Random Incomplete Graphs} (``RIG''): Given a desired number of vertices, $n$, and a desired number of edges, $n_e = \dim \mainmark{\mathbf{w}} < {n \choose 2}$, we randomly chose $n_e$ distinct pair of vertices to establish the edges between. Weights for the edges are sampled from an uniform distribution on the interval $[1-w_{p},1+w_{p}]$.
    \item[iii.] \emph{Social Network Graphs} (``Social''): As a representative of real-world networks, we extracted subgraphs from the ``Government'' graph category of the Gemsec Facebook Dataset \cite{rozemberczki2019gemsec} which encompasses various graphs representing blue verified Facebook page networks.
    To generate the subgraphs, ego graphs with a radius of $2$ were created. Nodes were randomly selected without replacement to serve as the center of each ego graph. Only the first $100$ subgraphs containing between $25$ and $200$ vertices that were generated were selected, resulting in a set of $100$ subgraphs with an average $109.82$ vertices and $867.69$ edges per subgraph.
\end{enumerate}
}

\subsubsection{Adversarial Force Sampling}
\changedNick{
\changedSB{For computing steady-state amplitudes for specific instances of simulation for a given graph, we need to sample the adversarial forcing vector, $\mathbf{f}$, and the adversarial forcing frequency, $\forcingomega$.

As described in Section~\ref{sec:stoch_model}, we assume that} the forcing vector $\mathbf{f}$ is sampled from an $(n-1)$-dimensional unit sphere. This is achieved by \changedSB{sampling} each element of the vector \changedSB{from} the standard normal distribution, \changedSB{and then normalizing the vector}~\cite{muller1959note}.
\changedAlp{Here we highlight that the necessary number of forcing vector samples to cover the sphere surface increases exponentially as the size of the network, $n$, (the dimension of the forcing vector/unit sphere) increases, if the sample dispersion is to be maintained. This comes from the fact that the dispersion is inversely proportional to the sample size and the dimension~\cite{Sukharev1971,Deheuvels1983}.}


\changedSB{As described in Section~\ref{sec:stoch_model}, the forcing frequency needs to be sampled using a probability density function that is a uniformly weighted sum of multiple Cauchy distributions each of which are centered at the natural frequencies, $\{\omega_j\}_{j=1,\dots,n}$, with a constant spread of $h$. However, in order to perform this sampling, one needs to compute the inverse of the cumulative distribution function (c.d.f.) of the $\rho$ described in \revAS{equation~\eqref{eq:pdf_omega}}, which is computationally difficult. We thus adopt \changedRBlum{a practical method that is used to generate samples from mixture models as explained in~\cite{Moitra_2018}:} Using an uniform probability of $1/n$ on all the natural frequencies, $\{\omega_j\}_{j=1,\dots,n}$, we first sample one of the natural frequencies, say $\omega_s$. Then the forcing frequency is sampled from a Cauchy distribution centred at $\omega_s$ and with a spread of $h$ using the inverse c.d.f. of Cauchy distribution, $\forcingomega = \omega_s + h \tan(\pi(p - 0.5)$, where $p$ is sampled from an uniform distribution over the unit interval $[0,1]$.
}



}

\begin{figure}
  \centering
  \includegraphics[scale=1]{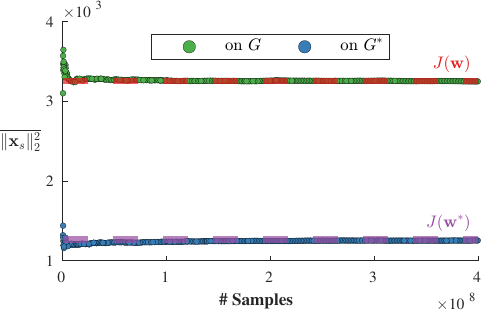}
  \caption{\changedAlp{Squared 2-norm of the steady state response $\mainmark{\mathbf{x}}_{s}$ is evaluated in closed-form using sampled external forcing. This plot shows the running average of $\|\mainmark{\mathbf{x}}_s\|^2_2$ against the number of forcing samples. The running averages are computed using batches of $100$K samples. Over large enough samples, the average $\overline{\|\mainmark{\mathbf{x}}_s\|^2_2}$ evaluated on the initial graph $\mainmark{G}$ and the optimized graph $\mainmark{G}^*$ converge to the values of the objective function evaluated at $\mainmark{\mathbf{w}}$ and $\mainmark{\mathbf{w}}^*$ respectively.}
  }
  \label{fig:cf_vs_obj_ori}
\end{figure}

\subsection{Network Graph Optimization}
First we present the results \changedSB{from Network Graph Optimization}.

\subsubsection{Validation of the Objective Function}
\label{sec:validation_obj_func}
The objective function, $\mainmark{J}$ \revAS{(Proposition~\ref{prop:network_vulnerability})}, for the network graph spectrum optimization problem is the expected value of the squared 2-norm of the steady-state response of the dynamic network subject to adversarial forcing with the stochastic model explained in Section~\ref{sec:stoch_model}.

To validate the the accuracy of the objective function in representing the expected value, we generate $400$M adversarial forcing samples (using the procedure explained in Section~\ref{sec:res-setup}), evaluate the closed-form steady state response, $\mainmark{\mathbf{x}}_s$, for each sample using \changedSB{equation}~\eqref{eq:ss_network} and compute the average squared 2-norm of the responses $\overline{\|\mainmark{\mathbf{x}}_s\|^2_2}$. For the validation study, we use a RCG with $n=10$ and $w_p=0.3$. The running average over the number of samples divided in multiple batches are provided in Figure~\ref{fig:cf_vs_obj_ori}.

It can be observed that over \changedRBlum{a} large number of forcing samples, the average of the squared 2-norm of the steady-state responses is well approximated by the objective values for both initial and optimized graphs. Hence, $\changedSB{\mathbb{E}_{\mathbf{f},\forcingomega}\left( \|\mathbf{x}_s\|^2_2 \right)}$ is accurately represented by $\mainmark{J}$. \changedSB{Consequently}, it can be seen that on the optimized graph, the steady state responses have smaller amplitudes on average. Following this validation, we can use the objective value as a measure of a graph's vulnerability to adversarial attacks, where a lower objective function indicates less vulnerability.

\begin{figure}
  \centering
  \includegraphics[scale=1]{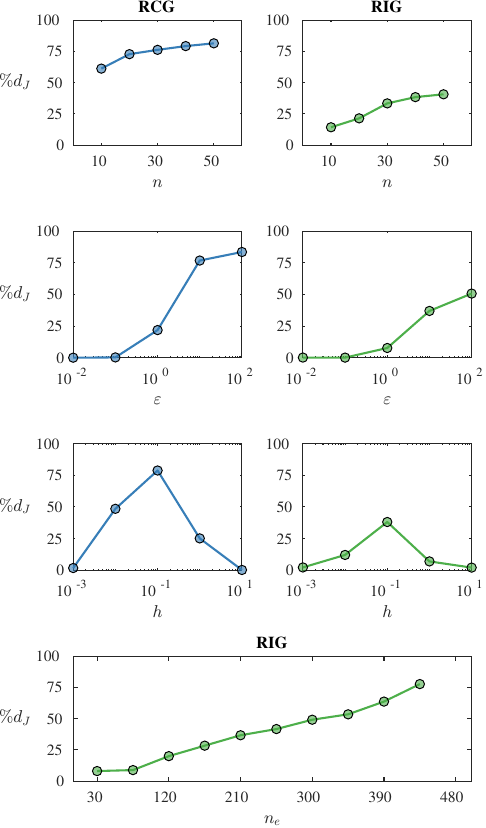}
  \caption{\changedAlp{A RCG and a RIG are generated for the problem instances specified by each set of parameters (only a RIG is generated for the case where the variable parameter is the number of edges). The optimization problem is solved for each instance, and the percentage decrease in objective values are plotted against the varying parameter.}}
  \label{fig:ori_pDecrease_param}
\end{figure}

\begin{figure}
  \centering
  \includegraphics{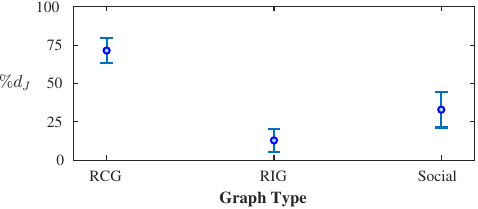}
  \caption{\changedAlp{The network graph spectrum optimization problem is solved for instances featuring 100 RCGs, 100 RIGs and 100 social network graphs. On problems instances where the main network graphs are complete, the optimization consistently yielded larger relative decrease in the objective values.}}
  \label{fig:ori_pDecrease_box}
\end{figure}




\subsubsection{Parameter Analysis}
Each spectrum optimization problem on a network graph can be specified via a set of parameters regarding the second order dynamics of the network, the external forcing and the constraints. These parameters are the number of vertices on the graph ($n$), the number of edges on the graph ($n_e$), the minimum weight constraint ($w^{min}$), the stiffness constant ($\varepsilon$), the damping factor ($\mainmark{\gamma}$), and the spread of the external agent's frequency distribution ($h$).

We analyze the effects of these parameters on the percentage reduction of objective value that can be achieved via the spectrum optimization, hence the reduction in the vulnerability of the main network graph using the \changedSB{Network Graph Optimization} method. For this purpose, we start with set of parameter values, $n=30$, $n_e = 225$, $w^{min}=10^{-3}$, $\varepsilon=10$, $\mainmark{\gamma} = 10^{-6}$, $h=0.1$, and generate problem instances featuring both RCGs and RIGs where we vary one parameter and keep the rest constant. We solve for each problem instance and compute the percentage reductions in objective as $\%d_{\mainmark{J}} = \frac{|\mainmark{J}^0 - \mainmark{J}^*|}{\mainmark{J}^0} \times 100$ (where superscripts $0$ and $*$ denote initial and optimal objective values), which are plotted against the varying parameter values in Figure~\ref{fig:ori_pDecrease_param}. Note that by comparing the percentage decrease in the objectives instead of the final objective values achieved, we are trying to isolate the effect of the parameters on the effectiveness of \changedSB{Network Graph Optimization} method in reducing the vulnerability of a graph instead of trying to find the set of problem parameters that make the network the least vulnerable.

It can be observed \changedSB{from Figure~\ref{fig:ori_pDecrease_param}} that a larger decrease in the objective value can be achieved as the number of vertices or the number of edges increase. Intuitively, more vertices and more edges correspond to more flexibility in distributing the weight resources, thus resulting in larger improvements to the vulnerability of the graph. As expected, larger stiffness yields better results, where the effect gets more significant with increased orders of magnitude. The spread of the external agent's frequency distribution has a non-monotonic effect. As the spread gets smaller, the agent is able to pick the resonance frequencies more accurately, leaving the graph helpless against the attack, whereas a larger frequency spread corresponds to an agent that almost arbitrarily picks its frequencies, against which any modification of the graph based on reasoning would be less effective. Since the minimum weight constraint and the damping factor did not demonstrate a significant effect on the percentage decrease of the objective, corresponding plots are excluded. By observing the plots overall and the analysis on the number of edges, it is clear that the spectrum optimization on a main network graph is more effective when the graph is complete. This behavior will become more apparent in the next section.

\subsubsection{Demonstration of the Effectiveness of \changedSB{Network Graph Optimization}}
To demonstrate the overall effectiveness of spectrum optimization on the main network graph in reducing the network vulnerability, we solve the optimization problem for RCGs, RIGs \changedSB{and \emph{Social} graphs,} and show that significant decrease in objective values can be achieved. We generate 100 RCGs and RIGs with $n$ sampled uniformly from the interval $[10,30]$ and $w_{p}$ sampled uniformly from the interval $[0.1,0.5]$. For the RIGs, we sampled $n_e$ from the interval $[n,n^2/4]$. The parameters associated with the network dynamics are the minimum weight, $w^{min}=0.001$, the stiffness constant, $\varepsilon=10$, the damping coefficient, $\mainmark{\gamma}=10^{-6}$, and the adversarial agent's frequency spread, $h=0.1$. The average percentage decrease in the objective value and the standard deviation across the problem instances featuring RCGs, RIGs and \changedSB{\emph{Social}} graphs are provided in Figure~\ref{fig:ori_pDecrease_box}.
\changedSB{Qualitatively, on the spectrum of the graph, the optimization is manifested as a \emph{flattening} of the spectrum, 
as can be seen for the
complete graph in Figure~\ref{fig:cg_ev_spectrum}
and the
\emph{Social} graph in Figure~\ref{fig:sg_ev_spectrum}.
}
%


\begin{figure*}

  \centering
  \includegraphics[scale=1]{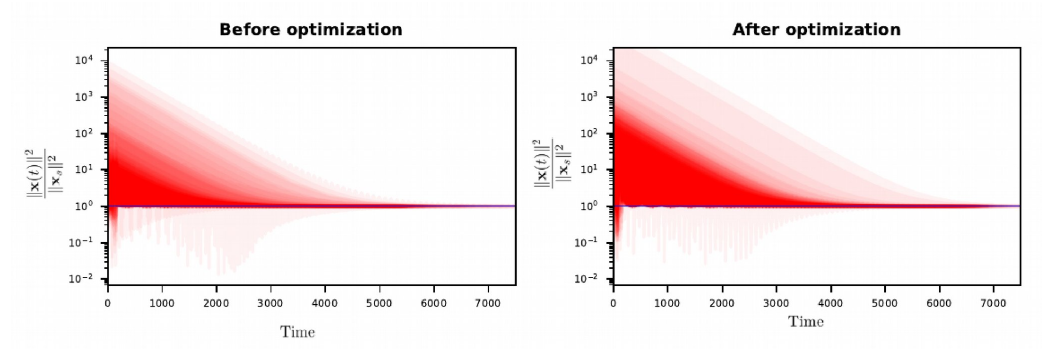}
   \vspace{-0.5em}
  \caption{\changedSB{Normalized amplitude} plot of 100 different \changedSB{numerical} simulations of the second-order dynamics for both the initial and optimized \changedSB{complete} graph, displaying their corresponding two-norm squared amplitude values divided by the steady-state squared amplitude value calculated by the closed-form evaluation for the respective forcing vector and forcing frequency \changedSB{(\emph{i.e.}, $\frac{\| \mathbf{x}(t)\|^2}{\|\mathbf{x}_s\|^2}$)}. Since the two-norm squared amplitude value from the simulation of the second-order dynamics should converge to the same closed-form steady-state evaluation for the two-norm squared amplitude given the same corresponding forcing vector and forcing frequency, the values in the plot are expected to approach 1, as indicated by the blue line, which they indeed do.
  \changedSB{In these $100$ simulations, the initial graph's mean steady-state squared amplitude value is 0.0899, while the optimized graph's mean steady-state squared amplitude value is 0.0145.}
}
  \label{fig:density_cg}

  \vspace{-1em}
\end{figure*}

As mentioned before, network graph spectrum optimization is more successful at reducing the objective value relative to the initial value of the objective when it is performed on complete graphs. A reason for this behavior is the greater vulnerability of the complete graphs to the resonance attacks, due to the fact that the natural frequencies of a complete graph \changedRBlum{are} heavily accumulated around a value \changedSB{resulting in a \emph{peaky} spectrum}, compared to a relatively \changedSB{flatter/uniform} distribution of the natural frequencies on an incomplete graph. 
\changedNick{
\changedSB{A fewer number of optimization variables impose greater rigidity on} incomplete graphs due to their fewer edges, whereas complete graphs, with their maximum possible number of edges, offer a greater \changedSB{flexibility in} edge weight manipulations.
\changedSB{Qualitatively,} this is \changedSB{manifested by a lower} relative \changedSB{flattening} of the spectrum 
\changedSB{in case of} the incomplete \changedSB{\emph{Social} graph (Figure~\ref{fig:sg_ev_spectrum}) as compared to the} complete graph \changedSB{(Figure~\ref{fig:cg_ev_spectrum})}.
An embedding of an \changedAlp{optimized \emph{Social}} graph is shown in Figure~\ref{fig:median-social}
}

\subsubsection{Numerical Second-Order Dynamics Simulation of the Main Network}
\label{sec:sim_explanation}
\changedNick{
\changedSB{We} \changedSB{simulated (performed numerical integration of)} the second-order dynamics \changedSB{on 100 different graphs, with} both the initial and optimized \changedSB{weights 
with} varying forcing vectors and sampled forcing frequencies. The simulations were run until a steady state was achieved 
\changedSB{and the final steady-state amplitude was noted.}

\emph{Complete Graph \changedSB{Numerical Simulations}:} 
\changedSB{We considered an unoptimized}
complete graph with \changedSB{uniform edge weights with an} added perturbation as detailed in Section ~\ref{sec:res-setup}, \changedSB{as well as the corresponding optimized graph obtained using the network graph optimization method detailed in Section~\ref{sec:network_graph_optimization}, and performed 100 numerical simulations on each of these graphs}.
\changedSB{The squared amplitude of $\mathbf{x}$ as a function of time for each of the 100 simulations, each normalized by the closed-form steady-state squared amplitude \changedAlp{$\|\mathbf{x}_s\|^2$}, is shown in Figure~\ref{fig:density_cg}. Besides observing that the steady-state amplitudes of the numerical simulations match the computed closed-form values, we note that the unsteady amplitude in relation to the steady-state amplitude has less variation in the optimized graph.}

\editsNick{
}}

\revAS{
\subsubsection{Application of Network Graph Optimization to a Robot Network}
We consider a team of $n$ mobile robots and their communication network described by a complete graph. The signal strength between robot $i$ and $j$ (represented by the edge weight $w_{ij}$) is computed as: $w_{ij} = \frac{A_{dist}}{\|r_i - r_j\| + \varepsilon_{dist}}$, where $A_{dist}$ and $\varepsilon_{dist}$ are some constants, and $r_i$, $r_j$ indicate the positions of robots $i$ and $j$.

We solve the optimization problem defined in Section~\ref{sec:eigenspectrum_optimization} and the constraints on the edge weights defined therein. In addition, we consider a physical constraint that prevents robot collision given as: \[\|r_i - r_j\| \geq d_{min} \quad \forall i,j \in \mathbb{Z} \text{ and } 1\leq i,j \leq n\] Then, the goal is to optimally relocate each robot such that the objective value is minimized.

We consider three types of initial configurations for the robots: arbitrary placement within some bounding box, on a uniform grid, on a line. We generate 10 instances for each initial condition where some small random perturbation is applied to the robot locations. Following parameters are used for the experiments: $n = 30$, $w^{min}=0.001$, $\varepsilon=1$, $\mainmark{\gamma}=10^{-6}$, $h=0.1$, $A_{dist} = 1$, $\varepsilon_{dist} = 0.1$, $d_{min} = 1$.

The mean and the standard deviation of the objective reduction achieved from each type of initial configuration is reported in Table~\ref{tab:rn-opt-table}. Initial and optimal robot locations for one instance of the problem are provided in Figure~\ref{fig:robot_network}.

\renewcommand{\arraystretch}{1.5} 
\begin{table}[h!]
	\centering
	\caption{Robot Network Optimization Results}
	\label{tab:rn-opt-table}
	\begin{tabular}{>{\columncolor[gray]{0.9}}c|c|c|c}
		\rowcolor[gray]{0.9} & \textbf{Arbitrary} & \textbf{Grid} & \textbf{Line} \\ \hline
		Mean $\%d_{\mainmark{J}}$ & $27.38$   & $39.20$ & $27.02$     \\ \hline
		Std. $\%d_{\mainmark{J}}$ & $4.05$    & $0.91$    & $0.64$  
	\end{tabular}
\end{table}

\begin{figure}[t!]
  \centering
  \includegraphics[scale=1]{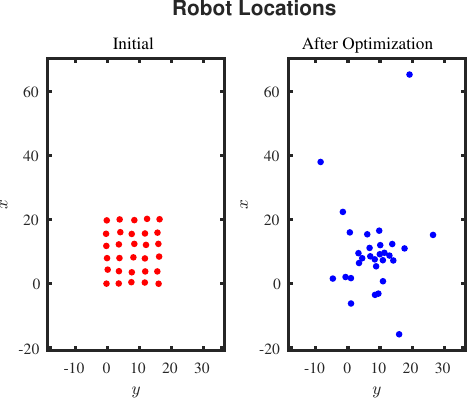}
  \caption{\revAS{Robots are initially arranged on a $6\times5$ grid with some small perturbations. After the optimization, robots are relocated to minimize the objective.}}
  \label{fig:robot_network}
\end{figure}
}

\subsection{Auxiliary Graph Optimization}
For the \changedSB{Auxiliary Graph Optimization} approach, we conduct similar experiments and provide additional analysis on the effects of auxiliary damping.

\subsubsection{Validation of the Objective Function}
The objective function $\auxmark{J}$ for the auxiliary graph spectrum optimization problem is the expected value of the squared 2-norm of the steady-state response corresponding to the main network vertices when the dynamic network is subject to stochastic adversarial forcing.
To validate the accuracy of the objective function in representing the expected value, we generate $800$M adversarial forcing samples, evaluate the closed-form steady state responses for each sample using equation~\eqref{eq:ss_cf_aux} and compute the average squared 2-norm of the responses. For the validation study, we generate a RCG with $n=10$, with $w_p = 0.3$ and use it as the main network graph. The running average over the number of samples divided in multiple batches are provided in Figure~\ref{fig:cf_vs_obj_aux_no_gap}.
\begin{figure}[t!]
  \centering
  \includegraphics[scale=1]{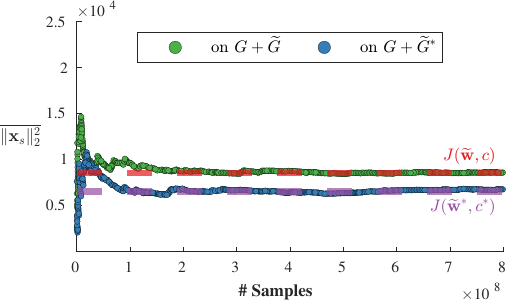}
  \caption{\changedAlp{Squared 2-norm of the steady state response corresponding to the main graph $\mainmark{\mathbf{x}}_s$ is evaluated in closed-form using sampled external forcing. This plot shows the running average of $\|\mainmark{\mathbf{x}}_s\|^2_2$ against the number of forcing samples. The running averages are computed using batches of $100$K samples. Over large enough samples, the average $\overline{\|\mainmark{\mathbf{x}}_s\|^2_2}$ evaluated on the unoptimized combined network system $\mainmark{G} + \auxmark{G}$ and the optimized system $\mainmark{G} + \auxmark{G}^*$ converge to the values of the objective function evaluated at $(\auxmark{\mathbf{w}},c)$ and $(\auxmark{\mathbf{w}}^*,c^*)$ respectively.}}
  \label{fig:cf_vs_obj_aux_no_gap}
\end{figure}

The problem instance generated for the validation study resulted in an optimized auxiliary graph for which $\mainmark{\Omega}$ and $\auxmark{\Omega}^*$ are simultaneously diagonalizable. As a consequence, we observe that $\overline{\|\mainmark{\mathbf{x}}_s\|^2_2}$ converge to the objective values for both the optimized and unoptimized combined networks. To demonstrate the fact that there will be an approximation error between $\overline{\|\mainmark{\mathbf{x}}_s\|^2_2}$ and $\auxmark{J}$, when $\mainmark{\Omega}$ and $\auxmark{\Omega}^*$ are not simultaneously diagonalizable, we perform another auxiliary graph spectrum optimization based on a RIG with $n=10$, $n_e = 25$ and $w_p=0.3$. The running average over the number of samples divided in multiple batches are provided in Figure~\ref{fig:cf_vs_obj_aux_gap}.

\begin{figure}[t!]
  \centering
  \includegraphics[scale=1]{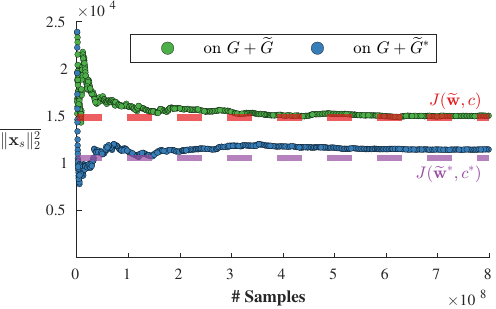}
  \caption{\changedAlp{Squared 2-norm of the steady state response corresponding to the main graph $\mainmark{\mathbf{x}}_s$ is evaluated in closed-form using sampled external forcing. This plot shows the running average of $\|\mainmark{\mathbf{x}}_s\|^2_2$ against the number of forcing samples. The running averages are computed using batches of $100$K samples. Over large enough samples, the average $\overline{\|\mainmark{\mathbf{x}}_s\|^2_2}$ evaluated on the unoptimized combined network system $\mainmark{G} + \auxmark{G}$ converges to the values of the objective function evaluated at $(\auxmark{\mathbf{w}},c)$. However, when evaluated on the optimized system $\mainmark{G} + \auxmark{G}^*$  the average $\overline{\|\mainmark{\mathbf{x}}_s\|^2_2}$ does not converge to the value of the objective function evaluated at $(\auxmark{\mathbf{w}}^*,c^*)$, resulting in an approximation error.}}
  \label{fig:cf_vs_obj_aux_gap}
\end{figure}

From Figures~\ref{fig:cf_vs_obj_aux_no_gap} and~\ref{fig:cf_vs_obj_aux_gap}, it can be observed that over \changedRBlum{a} large number of forcing samples, the average of the squared 2-norm of the steady-state responses is well approximated by the objective values when $\mainmark{\Omega}$ and $\auxmark{\Omega}$ are simultaneously diagonalizable, whereas there exist an approximation error when these matrices are not simultaneously diagonalizable. Also, it can be seen that on the optimized graphs, the steady state responses have smaller amplitudes on average.

As a sanity check, we leverage the theoretical result provided in equation~\eqref{eq:sanity} and confirm that both objective functions $\mainmark{J}$ and $\auxmark{J}$ match when evaluated numerically for arbitrary choices of $\mainmark{\Omega}$ and $\auxmark{\Omega}$ when $c=0$.

Following the validation of the objective function $\auxmark{J}$, we can use the objective value as a measure of a graph's vulnerability to adversarial attacks, where a lower objective function indicates less vulnerability.

\subsubsection{Parameter Analysis}
Parameters that specify an spectrum optimization problem on an auxiliary graph is similar to those of network graph optimization. Since the auxiliary graph edges and inter-graph edges are assumed to have non-negative weights, we do not consider the minimum weight constraint ($w^{min}$) parameter in this case. However, in addition to the network graph optimization parameters, we must consider the effects of the following parameters associated with auxiliary graphs: the auxiliary connectivity type (\emph{mirrored} or \emph{complete}), the weights resource multiplier $r_m$, and the auxiliary damping factor $\auxmark{\gamma}$. We defer the analysis of the auxiliary damping factor to Section~\ref{sec:res-aux-damp} and use a constant auxiliary damping factor of $\auxmark{\gamma} = 10^{-6}$ throughout the parameter analysis.

We analyze the effects of these parameters on the percentage reduction of objective value that can be achieved via the auxiliary graph optimization, hence the relative decrease in the vulnerability of the graph using the Auxiliary Graph Optimization method. We start with the same set of parameter values with the addition of $r_m = 5$, and generate problem instances where we vary one parameter and keep the rest constant. We solve for each problem instance and compute the percentage reductions in objective as $\%d_{\auxmark{J}} = \frac{|\mainmark{J}^0 - \auxmark{J}^*|}{\mainmark{J}^0} \times 100$, which are plotted against the varying parameter values in Figure~\ref{fig:aux_pDecrease_param}. Note that for the problem instances where the main network graph is a RIG, we provide two sets of results achieved with a mirrored auxiliary graph and a complete auxiliary graph.

Here we highlight that the percentage reduction of the objective value is computed based on the value of the objective before the auxiliary graph is attached, that is $\mainmark{J}^0$, instead of the objective value evaluated using an unoptimized auxiliary graph, that is $\auxmark{J}^0 = \auxmark{J}(\auxmark{\mathbf{w}},c)$. The individual effects of attaching an arbitrary auxiliary graph, and the optimization of the auxiliary graph will be presented in the next section.

\begin{figure}[t!]
  \centering
  \includegraphics[scale=1]{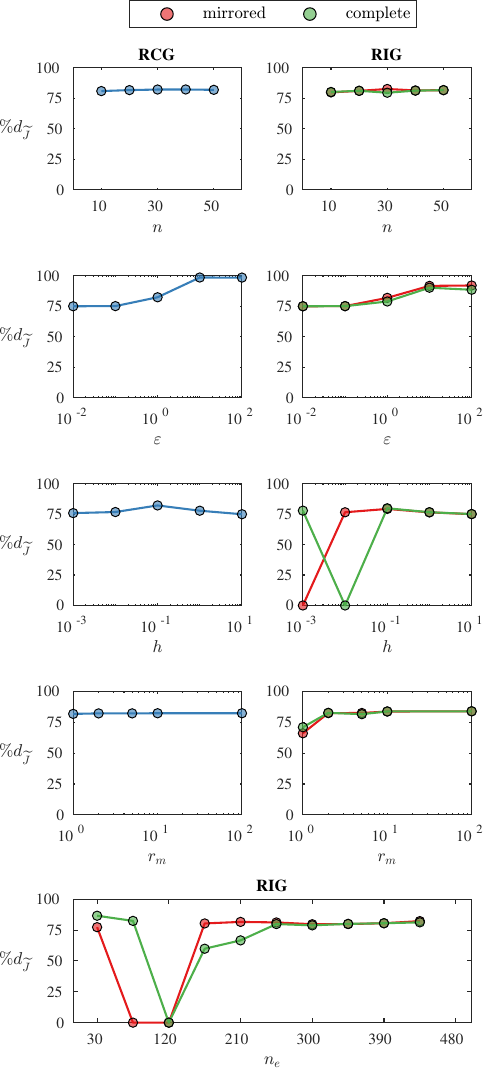}
  \caption{\changedAlp{A RCG and a RIG are generated for the problem instances specified by each set of parameters (only a RIG is generated for the case where the variable parameter is the number of edges). The optimization problem is solved for each instance (using both mirrored and complete auxiliary graphs for instances where the network graph is incomplete), and the percentage decrease in objective values ($\%d_{\auxmark{J}}$) are plotted against the varying parameter. Data points where the percentage decrease is at $0$ indicate the instances where the optimization failed to converge within the maximum number of iterations.}}
  \label{fig:aux_pDecrease_param}
\end{figure}

For all parameters, effects are similar to those on the network graph optimization. However, even for the parameter values for which the network graph optimization was less effective, \changedRBlum{the} Auxiliary Graph Optimization method can achieve larger decreases in the objective, which makes the approach less sensitive to the choice of the parameters. \changedRBlum{The s}ame insensitivity is observed to the weight resource multiplier parameter. For the instances where the network graph was incomplete, some of the optimizations of the mirrored auxiliary graph failed to converge in \changedRBlum{the} maximum number of iterations \changedRBlum{considered}, which is indicated by a $0\%$ decrease in the plots.

\subsubsection{Demonstration of the Effectiveness of Auxiliary Graph Optimization}
To demonstrate the overall effectiveness of spectrum optimization on the auxiliary graph in reducing the network vulnerability, we solve the optimization problem for RCGs and RIGs and show that significant decrease in objective values can be achieved. We use the same problem instances generated for the network graph optimization, with $r_m=5$ and $\auxmark{\gamma} = 10^{-6}$ and using complete auxiliary graphs. To demonstrate the effects of attaching an arbitrary auxiliary graph and the optimization of this auxiliary graph separately, we provide the average and the standard deviation of the percentage decrease in the objective calculated as (1) $\%d_{\auxmark{J}} = \frac{|\mainmark{J}^0 - \auxmark{J}^0|}{\mainmark{J}^0} \times 100$ (decrease achieved by going from network configuration $\mainmark{G}$ to $\mainmark{G} + \auxmark{G}$), and (2) $\%d_{\auxmark{J}} = \frac{|\mainmark{J}^0 - \auxmark{J}^*|}{\mainmark{J}^0} \times 100$ (decrease achieved by going from network configuration $\mainmark{G}$ to $\mainmark{G} + \auxmark{G}^*$) across the problem instances featuring complete and incomplete main network graphs are provided in Figure~\ref{fig:aux_pDecrease_box}. 

\begin{figure}[t!]
  \centering
  \includegraphics{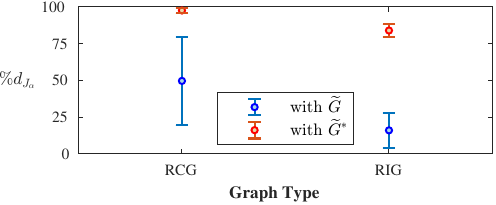}
  \caption{\changedAlp{The auxiliary graph spectrum optimization problem is solved for instances featuring 100 RCGs and 100 RIGs. We report the average and standard deviation of the percentage decrease in the objective achieved by both going from the network configuration $\mainmark{G}$ to $\mainmark{G} + \auxmark{G}$ and from the network configuration $\mainmark{G}$ to $\mainmark{G} + \auxmark{G}^*$. Success rates for running Auxiliary Graph Optimization on RCGs and RIGs were $\%100$ and $\%97$ respectively.}}
  \label{fig:aux_pDecrease_box}
\end{figure}

It can be seen that attaching even an arbitrary auxiliary graph decreases the vulnerability of the network significantly. However, performing the optimization over the auxiliary edge weights and inter-graph edges results in a further decrease of the vulnerability and provides more consistent behavior.

\subsubsection{Effect of the Auxiliary Damping and Auxiliary Damping Optimization} \label{sec:res-aux-damp}

Assuming that the auxiliary graph weights and the inter-graph edge weights are constant, the auxiliary objective function $\auxmark{J}$ becomes a function of the auxiliary damping factor $\auxmark{\gamma}$ only. Furthermore, if the auxiliary damping is uniform across all auxiliary vertices, $\auxmark{J}$ is a single-variable function. To visualize the effect of the auxiliary damping, we evaluate $\auxmark{J}$ on an optimized combined network (specified by $\mainmark{G}, \auxmark{G}^*, c^*$) with $\auxmark{\gamma}$ varying logarithmically on the interval $[10^{-6},10^5]$. The objective values are plotted against the auxiliary damping factor in Figure~\ref{fig:aux_obj_vs_damping}.

\begin{figure}[t!]
  \centering
  \includegraphics[scale=1]{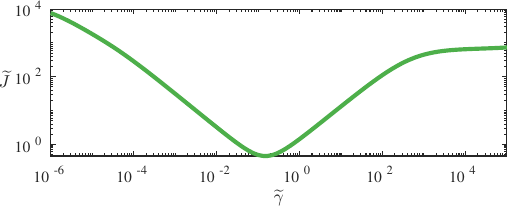}
  \caption{\changedAlp{Value of the auxiliary objective function $\auxmark{J}$ evaluated on an optimized combined network (specified by $\mainmark{G}, \auxmark{G}^*, c^*$) with auxiliary damping factor $\auxmark{\gamma}$ on the interval $[10^{-6},10^5]$.}}
  \label{fig:aux_obj_vs_damping}
\end{figure}

We observe that the objective function $\auxmark{J}$ is highly sensitive to the value of the auxiliary damping $\auxmark{\gamma}$ and that one can significantly decrease the objective value by setting the auxiliary damping to be larger than the damping on the main network. However, simply setting the auxiliary damping to the maximum allowed value does not yield the smallest objective value as observed from Figure~\ref{fig:aux_obj_vs_damping}. To the best of our understanding, as the auxiliary damping gets larger than the optimal value, the auxiliary network loses the ability to dissipate the signal that is being transmitted from the main network and the signal tends to bounce back causing a resonance. For this reason, optimizing over the variable $\auxmark{\gamma}$ could provide further improvements if the goal is to achieve the least possible vulnerability in the network.

}

\begin{figure*}
\begin{subfigure}
  \centering
  \includegraphics[scale=1.225]{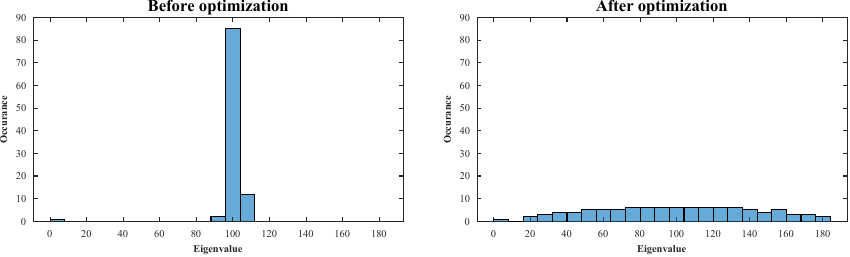}
  \caption{Example of eigenvalue spectrum of \changedSB{a} complete graph before and after \changedSB{optimization} represented as histograms of the eigenvalues of the networks stiffness matrix, $L+\varepsilon I$. The objective value significantly decreased from 1.378 to 0.3778, yielding a 72.58\% decrease as a result of the network graph optimization on a 100-node RCG. \changedSB{Qualitatively,} as a result of the network graph optimization, the eigenvalue spectrum has become \emph{flatter}.}
  \label{fig:cg_ev_spectrum}
  \end{subfigure}

  \begin{subfigure}
  \centering
  \includegraphics[scale=1.225]{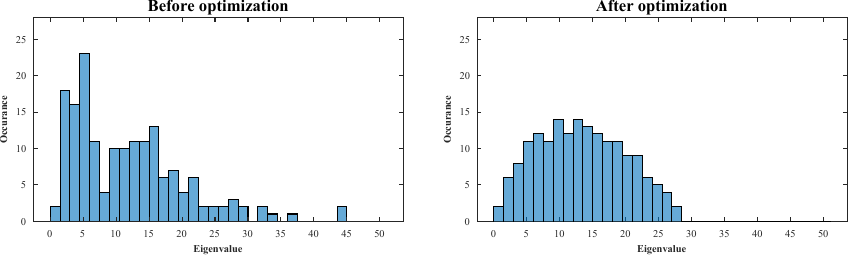}
  \caption{Example of eigenvalue spectrum of \changedAlp{a representative} Facebook Social subgraph before and after \changedSB{optimization} represented as histograms of the eigenvalues of the networks stiffness matrix, $L+\varepsilon I$. The objective value significantly decreased from 6.868 to 2.467 as a result of the network graph optimization on the 173-node Facebook social subgraph, corresponding to a 64.089\% decrease in the objective value. As a result of the optimization, the eigenvalue spectrum has become \changedSB{smoother and flatter}.}
  \label{fig:sg_ev_spectrum}
\end{subfigure}

\end{figure*}

\begin{figure}
  \centering
  \includegraphics[scale=1.2]{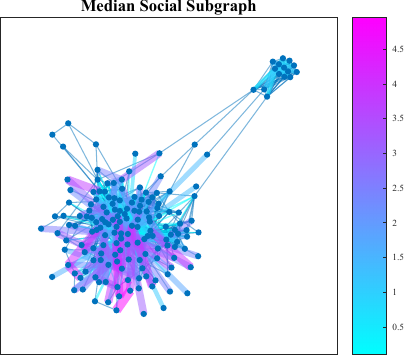}
  \caption{\changedSB{An embedding of an} \changedAlp{example} Facebook Social Subgraph with 173 \changedSB{vertices. The colors on the edges indicate weights after optimization.}} \label{fig:median-social}
\end{figure}

\section{Conclusion and Discussions}
\changedAlp{

In this paper, we developed the notion of vulnerability \changedSB{of a network with} second order \changedSB{signal} dynamics under adversarial forcing that obeys a known stochastic model. To minimize the network vulnerability, we proposed two methods that optimize the network structure: \emph{i.} The Network Graph Optimization method provides an optimal set of network edge weights under the condition that the edge weights can be directly manipulated, and, \emph{ii.} The Auxiliary Graph Optimization method allows us to design an auxiliary network that can be attached to the main network with the purpose of minimizing the vulnerability, when the main network edge weights cannot be adjusted directly. We conducted numerical experiments to analyze the two methods in detail.

Currently, the notion of vulnerability and the optimization problems posed in this work depend on \changedSB{a linear} model of the \changedSB{signal} dynamics and \changedSB{a specific stochastic model of} adversarial forcing.
\changedSB{While the adaptation of some aspects of the model to other setting (\emph{e.g.}, a different stochastic model of adversarial forcing) can be straight-forward re-derivation of the objective functions, a more general formulation that encompasses more complicated signal models, forcing models, and potentially nonlinear signal dynamics, is within the scope of future work.}
The optimization formulations presented in this paper lead to \changedSB{generally} non-convex problems which are in turn solved by gradient based solvers. While we do show convexity (Proposition~\ref{prop:convexity}) of the objective function of the Network Graph Optimization problem under the assumption that the parameter $h$ is large, a more general analysis of the optimization landscape for finite values of $h$ would be necessary to provide guarantees on the quality of the solution being returned, both for the Network Graph Optimization as well as the Auxiliary Graph Optimization problems. \changedSB{Such analyses are within the scope of future work}.

\changedSB{The current optimization problem is formulated as a centralized one that assumes complete knowledge of the network graph edge weights. A} potential future work involves the development of a distributed optimization scheme \changedSB{in which each vertex would use information about its local subgraph and would only adjusts weights on its incident edges in order to optimize the network}. A distributed method would allow the approach to scale to larger networks and generalize to settings where global information regarding the network may not be available due to privacy restrictions.
In future we will work towards implementing the proposed methods on real-world, physical networks such as electrical grids, robot networks and social networks.
}

\appendix
{\small\changedSB{
\subsection{Proof of Lemma~\ref{lemma:expected-value}} \label{ap:lemma1}

\emph{\textbf{Statement of the Lemma}
    If $\mathbf{f}\in\mathbb{R}^n$ is sampled from an uniform distribution over a $(n-1)$-unit sphere and $M$ is a symmetric matrix, then \[\mathbb{E}_{\mathbf{f}}(\|M \mathbf{f}\|_2^2) = \frac{1}{n} \|M\|_F^2 \]
    where $\|\cdot\|_F$ is the Frobenius norm.
    }

\begin{proof}
    Suppose $M$ is diagonalized by the orthogonal matrix, $U$, so that $M=U D U^T$, where $D = \mathrm{diag}(d_1,d_2,\cdots,d_n)$ is the diagonal matrix of the eigenvalues of $M$.
    
    Because of rotational symmetry of the distribution of $\mathbf{f}$ (uniform distribution over a sphere), the expected value of $\|M\mathbf{f}\|_2^2$ is independent of the choice of (an orthonormal) basis, and in particular, is the same in the basis of the eigenvectors of $M$. Thus, \begin{equation} \label{eq:EMf} \mathbb{E}_{\mathbf{f}}(\|M \mathbf{f}\|_2^2) = \mathbb{E}_{\mathbf{f}}(\|D \mathbf{f}\|_2^2) = \mathbb{E}_{\mathbf{f}}(\sum_{j=1}^n d_j^2 f_j^2) = \sum_{j=1}^n d_j^2 \,\mathbb{E}(f_j^2)\end{equation}
    where $\mathbb{E}(f_j^2)$ is the expected value of the square of the $j$-th component of $\mathbf{f}$.

    However, we note that because of the spherical symmetry of the distribution of $\mathbf{f}$, we must have $\mathbb{E}(f_1^2) = \mathbb{E}(f_2^2) = \cdots = \mathbb{E}(f_n^2) =: \xi$.
    Thus,
    \begin{eqnarray}
    & & \mathbb{E}_{\mathbf{f}}(\|\mathbf{f}\|_2^2) = 1 = \textstyle \sum_{j=1}^n \mathbb{E}(f_j^2) = n\xi \nonumber \\
    & \Rightarrow & \xi = 1/n \end{eqnarray} 
    Hence from \eqref{eq:EMf} we have, $\mathbb{E}_{\mathbf{f}}(\|M \mathbf{f}\|_2^2) = \sum_{j=1}^n d_j^2 /n = \frac{1}{n} \|M\|_F^2$.
\end{proof}

\subsection{Approximate Root Computation Using Linearization} \label{ap:root}

Consider a polynomial in the variable $x\in\mathbb{C}$, given by $Q(x, \gamma)$, where $\gamma\in\mathbb{R}$ is a parameter involved in the coefficients of the polynomial.
We are interested in approximately computing the roots of the polynomial for a general small, positive parameter value, $\gamma$, given the roots of the polynomial when $\gamma=0$ (which is presumed to be easier to compute).

If $\{r_k(\gamma)\}_{k=1,2,\cdots,n}$ are the roots of the polynomial $Q(x, \gamma)$ (possibly with multiplicity), we have
\begin{eqnarray*} 
& & Q(x,\gamma) ~=~ \prod_{k=1}^n \left( x-r_k(\gamma) \right) \\
& \Rightarrow & \frac{\partial Q}{\partial \gamma} (x,\gamma) ~=~ 
 -\sum_{l=1}^n r_l'(\gamma) \prod_{k\neq l} \left( x-r_k(\gamma) \right)
 \end{eqnarray*}
Evaluating the above at $x = r_j(\gamma)$,
 \begin{eqnarray*} 
 & & \frac{\partial Q}{\partial \gamma} (r_j(\gamma),\gamma) ~=~ -r_j'(\gamma) \prod_{k\neq j} \left( r_j(\gamma)-r_k(\gamma) \right) \\
  & \Rightarrow & r_j'(\gamma) ~=~ - \frac{ \frac{\partial Q}{\partial \gamma} (r_j(\gamma),\gamma) }{\prod_{k\neq j} \left( r_j(\gamma) - r_k(\gamma) \right)}
\end{eqnarray*}
This gives 1st order approximations for $r_j(\gamma)$ in the neighborhood of $\gamma=0$
\begin{eqnarray*} 
r_j(\gamma) & \approx & r_j(0) + r_j'(0) \gamma \\
            & = & r_j(0) - \frac{ \frac{\partial Q}{\partial \gamma} (r_j(0),0) }{\prod_{k\neq j} \left( r_j(0) - r_k(0) \right)} \gamma
\end{eqnarray*}

}}

\section*{Acknowledgements}

\changedSB{We gratefully acknowledge the support of AFOSR award number FA9550-23-1-0046.

We would like to thank Dr.  Brian M. Sadler, Senior Research Fellow, University of Texas at Austin, for his valuable insights and discussions on the motivation and potential applications of this work during the course of writing this paper.}

\bibliographystyle{IEEEtran}
\bibliography{references}

\end{document}